\documentclass{article}

\usepackage{microtype}
\usepackage{graphicx}
\usepackage{subcaption}
\usepackage{booktabs} 

\usepackage{hyperref}

\usepackage[preprint]{icml2026}

\usepackage{amsmath}
\usepackage{amssymb}
\usepackage{mathtools}
\usepackage{amsthm}

\usepackage{booktabs}
\usepackage{multirow}

\usepackage{epigraph} 

\usepackage{listings}
\usepackage{xcolor}

\lstdefinestyle{promptstyle}{
  basicstyle=\ttfamily\small,   
  breaklines=true,              
  frame=single,                 
  backgroundcolor=\color{gray!10}, 
  xleftmargin=2mm,              
  xrightmargin=2mm,             
  numbers=none,                 
  columns=fullflexible           
}

\usepackage[most]{tcolorbox}

\newtcolorbox{ExampleBox}{
    colback=gray!10,      
    colframe=gray!10,     
    arc=3pt,              
    outer arc=3pt,        
    boxrule=0pt,          
    left=5pt,             
    right=5pt,            
    top=5pt,              
    bottom=5pt,           
    width=\linewidth,     
    enhanced,              
    drop fuzzy shadow,    
    shadow={0.1mm}{-0.1mm}{0.2mm}{black!10},  
}

\usepackage{etoolbox}

\AtBeginDocument{
  \setlength{\abovedisplayskip}{6pt}       
  \setlength{\belowdisplayskip}{6pt}       
  \setlength{\abovedisplayshortskip}{6pt}  
  \setlength{\belowdisplayshortskip}{6pt}  
}

\usepackage{caption}
\usepackage[capitalize,noabbrev]{cleveref}

\theoremstyle{plain}
\newtheorem{theorem}{Theorem}

\theoremstyle{definition}
\newtheorem{definition}[theorem]{Definition}

\theoremstyle{remark}

\usepackage[textsize=tiny]{todonotes}

\icmltitlerunning{\hfill}

\begin{document}

\twocolumn[
  \icmltitle{Interpreting Emergent Extreme Events in Multi-Agent Systems}



\icmlsetsymbol{equal}{*}  
\icmlsetsymbol{corr}{$\dagger$}

\begin{icmlauthorlist}
  \icmlauthor{Ling Tang}{ailab,sjtu,equal}   
  \icmlauthor{Jilin Mei}{ailab,fdu,equal}    
  \icmlauthor{Dongrui Liu}{ailab,corr}       
  \icmlauthor{Chen Qian}{ailab,ren}       
  \icmlauthor{Dawei Cheng}{tj}       
  \icmlauthor{Jing Shao}{ailab}       
  \icmlauthor{Xia Hu}{ailab}       
\end{icmlauthorlist}

\icmlaffiliation{ailab}{Shanghai Artificial Intelligence Laboratory}
\icmlaffiliation{sjtu}{Shanghai Jiao Tong University}
\icmlaffiliation{fdu}{Fu Dan University}
\icmlaffiliation{tj}{Tongji University}
\icmlaffiliation{ren}{Renmin University of China}

\icmlcorrespondingauthor{Dongrui Liu}{liudongrui@pjlab.org.cn}

  \icmlkeywords{Machine Learning, ICML}

  \vskip 0.3in
]



\printAffiliationsAndNotice{\icmlEqualContribution}

\begin{abstract}
Large language model-powered multi-agent systems have emerged as powerful tools for simulating complex human-like systems. The interactions within these systems often lead to extreme events whose origins remain obscured by the black box of emergence. Interpreting these events is critical for system safety. This paper proposes the first framework for explaining emergent extreme events in multi-agent systems, aiming to answer three fundamental questions: \textit{When does the event originate? Who drives it? And what behaviors contribute to it?} Specifically, we adapt the Shapley value to faithfully attribute the occurrence of extreme events to each action taken by agents at different time steps, \emph{i.e.}, assigning an attribution score to the action to measure its influence on the event. We then aggregate the attribution scores along the dimensions of time, agent, and behavior to quantify the risk contribution of each dimension. Finally, we design a set of metrics based on these contribution scores to characterize the features of extreme events. Experiments across diverse multi-agent system scenarios (economic, financial, and social) demonstrate the effectiveness of our framework and provide general insights into the emergence of extreme phenomena. The source code is available at \url{https://github.com/mjl0613ddm/IEEE}.
\end{abstract}

\section{Introduction}

\begin{figure*}[ht]
  \begin{center}
  \vskip 0.2in

    \centerline{\includegraphics[width=\linewidth]{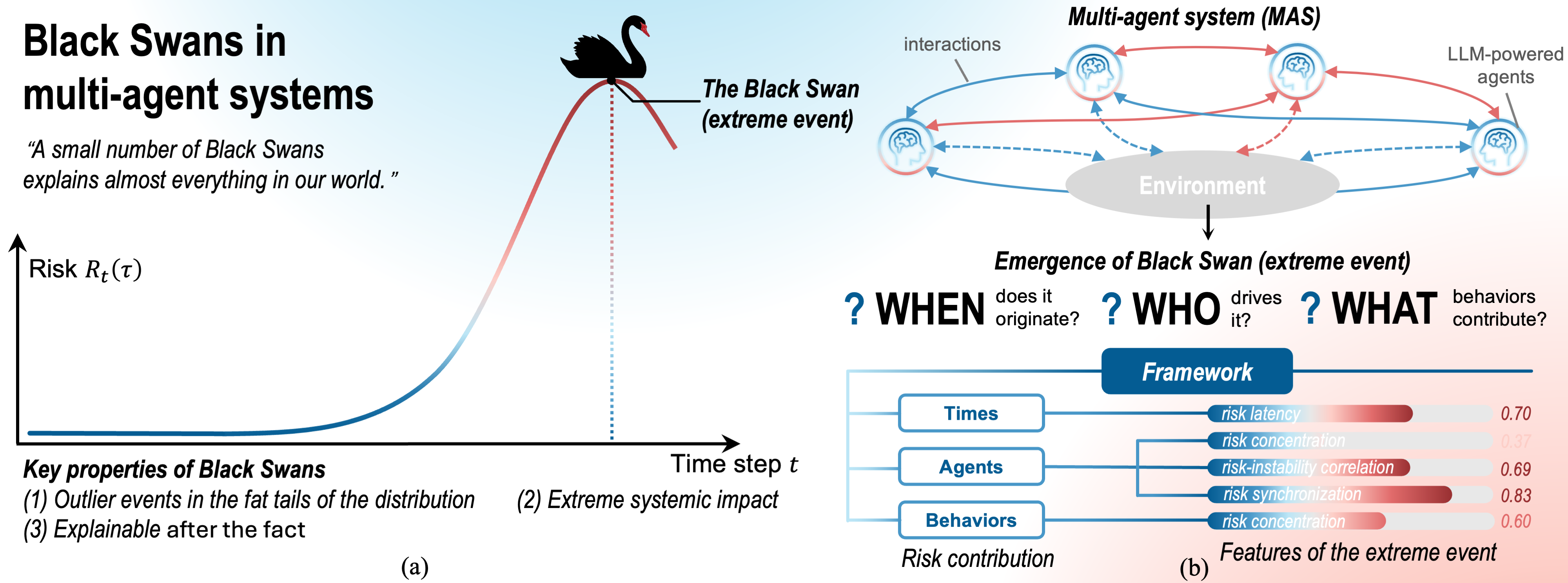}}
    \caption{Overview of the proposed framework for explaining Black Swans (emergent extreme events) in MAS.  (a) Black Swans are defined as statistical outliers in the fat tails of risk distributions, carrying extreme systemic impact and being explainable only after the occurrence.  (b) We explain extreme events by answering the questions from the perspective of \textit{when}, \textit{who}, and \textit{what}. Specifically, we use attribution methods to quantify the risk contributions across these three dimensions. Based on these contributions, we design a set of metrics to quantitatively analyze the features of extreme events.}
    \label{fig:main}
  \end{center}
  \vskip -15pt
\end{figure*}

In recent years, large language models (LLMs) have exhibited their remarkable capacity to capture human behavioral patterns in language, decision-making, and social interactions~\cite{grossmann2023ai, xi2025rise}. These capabilities have made it possible to build highly human-like multi-agent systems (MAS) that consist of multiple LLM-powered agents. Recent  studies have increasingly leveraged LLM-powered MAS to simulate complex systems~\cite{park2023generative,yang2024oasis,li2024econagent,yang2025twinmarket,wang2025decoding}. In such simulations, LLM-powered agents interact autonomously with each other and the environment, leading to the emergence of various complex phenomena.

\begin{quote} \vspace{-5pt}
    \small \itshape
    ``A small number of Black Swans explains almost everything in our world.''
    \par\hfill ---  \textit{The Black Swan}~\cite{nicholas2008black}
\end{quote}\vspace{-5pt}

Among the phenomena emerging from MAS, we focus on \textbf{\textit{Black Swans}}~\cite{nicholas2008black}. As illustrated in Figure~\ref{fig:main}(a), Black Swans refer to extreme events characterized by three key properties: (1) they are \textit{outliers} falling outside regular expectations, typically residing in the tail of a fat-tailed distribution~\cite{clauset2009power,taleb2020statistical}; (2) they carry an \textit{extreme impact}, \emph{e.g.}, causing systemic collapse or drastic state transitions; and (3) they feature \textit{retrospective predictability}, meaning their occurrence is interpretable only after the fact.  Existing literature has validated the emergence of such events in MAS, including soaring inflation in economic systems~\cite{li2024econagent}, market crashes in financial markets~\cite{yang2025twinmarket}, and group polarization in social networks~\cite{wang2025decoding}.

However, interpreting emergent extreme events in MAS remains a critical challenge. The complex interactions between agents and their environment create a black box of emergence~\cite{Miller2017ExplanationIA,baker2020emergenttoolusemultiagent}. The key factors across different dimensions that lead to the disaster remain hidden, leaving fundamental questions underexplored: \textbf{\textit{When does the extreme event originate? Who drives it? And what behaviors contribute to it?}} Explaining the extreme events not only provides a theoretical analysis of the underlying mechanisms but also offers actionable guidance for preventing disasters, \emph{e.g.}, restricting high-risk agents or behaviors. However, there is no existing framework to explain such emergent extreme events in LLM-powered MAS to the best of our knowledge.

\begin{figure*}[ht]
  \begin{center}
  \vskip 0.2in

    \centerline{\includegraphics[width=\linewidth]{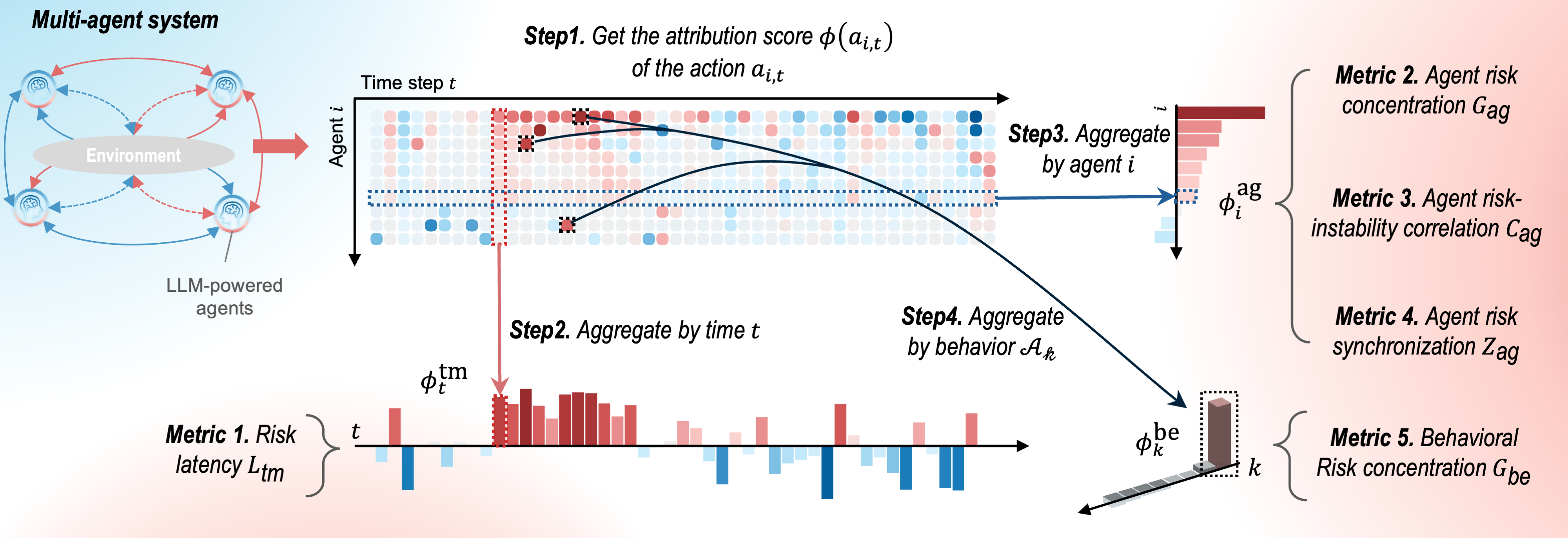}}
    \caption{Risk attribution process and quantitative metric derivation. \textbf{Step 1: action-level attribution.} We first attribute the extreme event to all individual actions $a_{i,t}$ in the trajectory leading to the event, assigning each an attribution score $\phi(a_{i,t})$. Red indicates actions that increase the risk of the extreme event, while blue represents actions that decrease it. \textbf{Step 2-4: dimensional aggregation.} These attribution scores are subsequently aggregated along three dimensions of time $\phi^{\text{tm}}_{t}$, agent $\phi^{\text{ag}}_{i}$, and behavior $\phi^{\text{be}}_{k}$ to quantify their respective contributions. \textbf{Metrics.} Based on these aggregated contributions, we derive five quantitative metrics to characterize the extreme event: relative risk latency $L_{\text{tm}}$, agent risk concentration $G_{\text{ag}}$, risk-instability correlation $C_{\text{ag}}$, agent risk synchronization $Z_{\text{ag}}$, and behavioral risk concentration $G_{\text{be}}$.}
    \label{fig:metric}
  \end{center}
\vskip -22pt

\end{figure*}

In this paper, we propose the first framework to explain emergent extreme events in MAS. As shown in Figure~\ref{fig:main}(b), this framework allows us to answer the questions from the perspectives of \textit{when}, \textit{who}, and \textit{what}. We aim to identify the key temporal phases, agents, and behaviors that contribute most to the extreme events. Specifically, as Figure~\ref{fig:metric} shows, we assign an attribution score to each agent's action to quantify its contribution to the extreme event based on the Shapley value~\cite{shapley1953value}.  Then, we aggregate the action-level attribution scores across time steps, agents, and types of behaviors to quantify the contributions of each dimension. Furthermore, we design a set of metrics based on these aggregated contributions to characterize the features of the extreme events.

We applied our framework to three distinct representative MAS scenarios: economic systems, financial markets, and social networks. For instance, in financial markets simulations, we find extreme volatility originates from immediate shocks. A few agents using short-term strategies drive these events, while the behavior of trading specific indices contributes the majority of the risk. More broadly, our analysis across diverse scenarios yields key insights into extreme events in MAS, which are summarized as follows.  

\begin{ExampleBox}
{ \itshape
$\bullet$ Insight 1. Extreme events originate from either early dormant risks or immediate shocks. 

$\bullet$ Insight 2. Extreme events are typically driven by a small subset of agents.

$\bullet$ Insight 3. Agents with high risk contribution often exhibit high instability. 

$\bullet$ Insight 4. Agents tend to increase or decrease risk synchronously. 

$\bullet$ Insight 5. A small number of behavior patterns contribute the majority of the risk leading to extreme events. }\end{ExampleBox}

\section{Related Work}

\textbf{LLM-based MAS for simulations:} Since LLMs demonstrate the capability to capture human reasoning and complex social interaction behaviors~\cite{wei2022chain,park2023generative,zhao2024competeai}, multi-agent systems (MAS) composed of these models are widely used for human-like simulations~\cite{park2023generative,li2023camel,yang2024oasis,li2024econagent,yang2025twinmarket,wang2025decoding}. Recently, LLM-based multi-agent simulations have moved beyond normal distribution scenarios, and they increasingly reveal extreme events in the fat tail of the distribution. For instance, \citet{li2024econagent} simulated the quick rise of inflation, \citet{yang2025twinmarket} reproduced market crashes, and \citet{wang2025decoding}  replicated extreme echo chamber effects in social networks. However, existing studies are primarily limited to validating the existence of these events, rather than interpreting them. 

\textbf{Attribution in MAS:} Extreme events observed in complex MAS usually stem from highly non-linear emergence~\cite{holland2000emergence,sornette2009dragon,park2023generative}. Attribution methods provide a way to interpret black-box models~\cite{zhou2016learning,lundberg2017unified,sundararajan2017axiomatic,wang2022self,zheng2023judging}. They quantify the marginal contribution of inputs to the system's output. Recently, these techniques have been increasingly adapted to MAS, aiming to quantify how agent actions influence the systemic outcomes. For instance, \citet{ge2025introducing} executed the system multiple times to statistically aggregate high-frequency behavioral patterns for identifying dangerous actions. \citet{zhang2025agentracer} trained a surrogate model to output attribution scores, and \citet{zhang2025agent, cemri2025multi} directly queried an LLM via prompting to derive attributions. These attribution methods for MAS depend on statistical priors or external knowledge, limiting their performance on the extreme events in the fat tail of the distribution. In comparison, we propose a faithful, prior-free attribution method to interpret these extreme events.

\textbf{Quantitative analysis of extreme events: } To quantitatively analyze extreme events, previous works designed metrics from different aspects. For example, \citet{gabaix2009power, sornette2017stock} used log-periodic patterns to show the extent of risk development, \citet{scheffer2009early,dakos2012methods} used autocorrelation to gauge the likelihood of occurrence, and \citet{li2023camel} used the Gini coefficient to measure the severity of extreme events. While these metrics can detect the occurrence of extreme events, they can not explain them. In comparison, we design a set of metrics based on faithful attribution scores to interpret the extreme events by answering the questions of \textit{when}, \textit{who}, and \textit{what}.

\begin{table*}[!t]
    \centering
    \caption{Experimental results on the approximation accuracy of Monte Carlo sampling. We report the cosine similarity between the ground truth value and the approximated value $\textbf{cos}(\textbf{vec}(\hat{\Phi}), \textbf{vec}(\Phi^*))$ with different sample sizes $M$. Results are reported with the mean and standard deviation across independent runs. GPT, Llama, Claude, Qwen, and DS denote the model GPT-4o mini, Llama-3.1-8B-Instruct, Claude-3-Haiku, Qwen-Plus, and DeepSeek-V3.2, respectively.}
    \label{tab:mc_accuracy}
    
    \resizebox{\linewidth}{!}{%
        \begin{tabular}{lcccccccc}
            \toprule
            \multirow{2}{*}{\textbf{Model}} & \multicolumn{4}{c}{\textbf{EconAgent}} & \multicolumn{4}{c}{\textbf{TwinMarket}} \\
            \cmidrule(lr){2-5} \cmidrule(lr){6-9}
             & $M=10^1$ & $M=10^2$ & $M=10^3$ & $M=10^4$ & $M=10^1$ & $M=10^2$ & $M=10^3$ & $M=10^4$ \\
            \midrule
            
            GPT    & $0.931 \pm 0.078$ & $0.973 \pm 0.042$ & $0.991 \pm 0.013$ & $0.995 \pm 0.008$ 
                   & $0.915 \pm 0.095$ & $0.995 \pm 0.003$ & $0.999 \pm 0.001$ & $1.000 \pm 0.000$ \\
            
            Llama  & $0.942 \pm 0.087$ & $0.972 \pm 0.048$ & $0.993 \pm 0.009$ & $0.997 \pm 0.003$ 
                   & $0.663 \pm 0.161$ & $0.950 \pm 0.046$ & $0.995 \pm 0.002$ & $0.999 \pm 0.001$ \\
            
            Claude & $0.976 \pm 0.033$ & $0.983 \pm 0.025$ & $0.996 \pm 0.005$ & $0.998 \pm 0.002$ 
                   & $0.969 \pm 0.027$ & $0.996 \pm 0.004$ & $0.999 \pm 0.001$ & $1.000 \pm 0.000$ \\

            Qwen   & $0.962 \pm 0.032$ & $0.986 \pm 0.011$ & $0.997 \pm 0.001$ & $0.999 \pm 0.001$ 
                   & $0.714 \pm 0.171$ & $0.975 \pm 0.020$ & $0.997 \pm 0.001$ & $0.999 \pm 0.001$ \\
                                        
            DS     & $0.944 \pm 0.052$ & $0.986 \pm 0.009$ & $0.996 \pm 0.001$ & $0.999 \pm 0.001$ 
                   & $0.839 \pm 0.072$ & $0.978 \pm 0.009$ & $0.998 \pm 0.001$ & $1.000 \pm 0.000$ \\
            \bottomrule
        \end{tabular}%
    }
    \vskip -10pt
\end{table*}

\section{Method}

This section presents a framework to analyze extreme events in MAS. We aim to answer three key questions: \textit{When does the extreme event originate? Who drives it? And what behaviors contribute to it?} Specifically, we formulate the problem of extreme event attribution and use the Shapley value to compute faithful attribution scores. We then aggregate these scores along the dimensions of time, agents, and behaviors. Finally, we design interpretable metrics to characterize the features of the extreme event.

\subsection{Preliminaries: Extreme Event Attribution in MAS}\label{subsec:pre}

We model the LLM-powered MAS as a tuple $\langle \mathcal{N}, \mathcal{S}, \mathcal{A}, P \rangle$. Let $\mathcal{N} = \{1, \dots, N\}$ denote the set of $N$ agents. $\mathcal{S}$ and $\mathcal{A}$ represent the global state space and the action space, respectively. At each discrete time step $t$, the system is in a global state $s_t \in \mathcal{S}$. Simultaneously, each agent $i$ executes an action $a_{i,t} \in \mathcal{A}$. The transition function $P$ specifies the probability distribution of the next state $s_{t+1}$, given the current state $s_t$ and joint actions of all agents $\{a_{1,t}, \dots, a_{N,t}\}$.  Consequently, the interaction process generates a trajectory $\tau = (s_1, \{a_{1,1},a_{2,1}\dots\}, s_2, \dots, s_T)$, where $T$ denotes the final time step.

\textbf{Problem Formulation.}  To quantify the extreme event within the trajectory $\tau$, we introduce a set of real-valued functions $\{R_t(\tau)|t=1,\dots,T\}$ as the metrics to measure the event risk at step $t$. As Figure~\ref{fig:main}(a) shows, an extreme event is defined to occur when the risk metric $R_t(\tau)$ exceeds a threshold\footnote{The threshold $\rho$ is set by domain experts. Please refer to Appendix \ref{app:scenario} for the definition of the specific system.} $\rho$.  For convenience, we set $T$ as the first time when the extreme event occurs, \emph{i.e.},  $R_T(\tau) > \rho$ while $\forall t < T,\; R_t(\tau) \le \rho $.  Our primary objective is to decompose the final risk $R_T(\tau)$ by assigning an attribution score $\phi(a_{i,t})$ to each action $a_{i,t}$ taken by agent $i$ at time $t$. Specifically, $\phi(a_{i,t})$ quantifies the marginal contribution of action $a_{i,t}$ to the occurrence of the extreme event. This decomposition is formulated as:
 \begin{equation}
    R_T(\tau) = \sum_{t=1}^{T} \sum_{i = 1}^{N} \phi(a_{i,t}).
\end{equation}

\textbf{Attributing the extreme event using Shapley values.} The Shapley value was originally introduced in game theory~\cite{shapley1953value}. Consider a game with multiple players, where each player aims to optimize their returns. The Shapley value is widely recognized as a unique, unbiased approach for fairly allocating the total reward to each player. The allocation \textbf{satisfies the axioms of \textit{efficiency}, \textit{symmetry}, \textit{nullity}, and \textit{linearity}}~\cite{roth1988shapley}. Please refer to Appendix~\ref{app:shapley} for the detailed proof.

We use the Shapley value to calculate the faithful attribution score $\phi(a_{i,t})$. As shown in Figure~\ref{fig:metric}, given the trajectory $\tau$ and the final risk $R_T(\tau)$, we define the set of players as the collection of all executed actions $\Omega = \{a_{1,1}, \dots, a_{N,T}\}$, where $|\Omega| = NT$. We consider the occurrence of the extreme event in the MAS as a game $v$. Let $2^{\Omega} = \{S \mid S \subseteq \Omega \}$ denote all possible subsets of $\Omega$. Let $\tau^{S}$ denote a counterfactual trajectory where actions within $S$ are preserved, while actions in $\Omega \setminus S$ are removed\footnote{Specifically, the counterfactual trajectory is generated via re-simulation. Actions in $\Omega \setminus S$ are replaced with expert-annotated safe actions $\tilde{a}_{i,t}$, while actions in $S$ are forced to replay. The system states $s_t$ evolve naturally following the standard system dynamics.}. The characteristic function $v(S)$ of the game is defined as the risk of the modified trajectory as $v(S) = R_T(\tau^S)$. The contribution $\phi(a_{i,t})$ of each action $a_{i,t}$ is then calculated using the Shapley value as follows:
\begin{equation}\label{eq:shapley} 
\phi(a_{i,t}) = \sum_{S \subseteq \Omega \setminus \{a_{i,t}\}} w_{|S|} \cdot \left( v(S \cup \{a_{i,t}\}) - v(S) \right),
\end{equation}
where the weight $w_{|S|} = |S|! (|\Omega| - |S| - 1)!/|\Omega|!$. 

\textbf{Estimating Shapley values using Monte Carlo Sampling}. Computing exact Shapley values entails an exponential time complexity of $O(2^{|\Omega|})$ as it requires evaluating all possible subsets of actions. To ensure computational feasibility, we follow~\cite{grabisch1999axiomatic} to employ Monte Carlo Sampling, which reduces the complexity to $O(M \cdot |\Omega|)$ given a sample size of $M$. Please refer to Appendix~\ref{app:mc} for details of the algorithm.

\textbf{Verifying the approximation accuracy of the Monte Carlo Sampling}.  Let $\Phi^* = [\phi^*(a_{i,t})] \in \mathbb{R}^{N \times T}$ denote the exact attribution matrix, where each entry $\phi^*(a_{i,t})$ corresponds to the exact Shapley value calculated in Equation~\ref{eq:shapley}. Similarly, let $\hat{\Phi} = [\hat{\phi}(a_{i,t})] \in \mathbb{R}^{N \times T}$ denote the estimated attribution matrix, where $\hat{\phi}(a_{i,t})$ represents the approximated value derived from Monte Carlo Sampling. We used the cosine similarity\footnote{The cosine similarity between two vectors lies in the range $[-1,1]$, where larger values indicate higher similarity.} $\textbf{cos}(\textbf{vec}(\hat{\Phi}), \textbf{vec}(\Phi^*))$  to measure the approximation accuracy, where $\textbf{vec}(\cdot)$ denotes the vectorization of a matrix. We conducted experiments using representative LLMs, including GPT-4o mini~\cite{hurst2024gpt}, Llama-3.1-8B-Instruct~\cite{grattafiori2024llama}, Claude-3-Haiku~\cite{anthropic2024claude}, 
Qwen-Plus~\cite{yang2025qwen3} and DeepSeek-V3.2~\cite{liu2024deepseek} on two different scenarios: EconAgent~\cite{li2024econagent} and TwinMarket~\cite{yang2025twinmarket}. Please refer to Section~\ref{subsec:setup} for detailed settings of different scenarios. We set the agent number $N=4$ and the number of time steps $T = 5$. For each experimental setting, we report the average results over 10 independent runs with different random seeds.  As shown in Table~\ref{tab:mc_accuracy}, the cosine similarity increases with the sample size. Specifically, with a sample size of $M=10^3$, the cosine similarity typically exceeds $0.99$. To balance efficiency and accuracy, we adopt $M=10^3$ in this paper unless otherwise specified.

\begin{table*}[!t]
    \centering
    \caption{Experimental results for quantitatively analyzing extreme events.  We report metrics to characterize the features of extreme events across three dimensions: 
\textit{When} (relative risk latency $L_{\text{tm}}$), 
\textit{Who} (agent risk concentration $G_{\text{ag}}$, risk-instability correlation $C_{\text{ag}}$, agent risk synchronization $Z_{\text{ag}}$), 
and \textit{What} (behavior risk concentration $G_{\text{be}}$). Results are reported with the mean and standard deviation across independent runs. }
    \label{tab:metric}
    \resizebox{0.97\linewidth}{!}{%
    \begin{tabular}{l @{\hspace{0.3cm}} l @{\hspace{0.5cm}} c ccc c}
        \toprule
        \multirow{2}{*}{\textbf{Scenario}} & \multirow{2}{*}{\textbf{Model}} 
        & \textbf{When} 
        & \multicolumn{3}{c}{\textbf{Who}} 
        & \textbf{What} \\
        \cmidrule(lr){3-3} \cmidrule(lr){4-6} \cmidrule(lr){7-7}
         & & $L_{\text{tm}}$ & $G_{\text{ag}}$ & $C_{\text{ag}}$ & $Z_{\text{ag}}$ & $G_{\text{be}}$ \\
        \midrule
        
        \multirow{5}{*}{\textbf{EconAgent}} 
         & GPT-4o mini    & $0.702 \pm 0.188$ & $0.516 \pm 0.100$ & $0.805 \pm 0.193$ & $0.505 \pm 0.117$ & $0.738 \pm 0.062$ \\
         & Llama-3.1-8B-Instruct  & $0.691 \pm 0.255$ & $0.610 \pm 0.137$ & $0.853 \pm 0.166$ & $0.400 \pm 0.183$ & $0.673 \pm 0.070$ \\
         & Claude-3-Haiku & $0.464 \pm 0.294$ & $0.459 \pm 0.086$ & $0.684 \pm 0.270$ & $0.597 \pm 0.141$ & $0.627 \pm 0.053$ \\
         & Qwen-Plus   & $0.660 \pm 0.278$ & $0.583 \pm 0.039$ & $0.827 \pm 0.242$ & $0.368 \pm 0.139$ & $0.513 \pm 0.122$ \\
         & DeepSeek-V3.2     & $0.794 \pm 0.116$ & $0.519 \pm 0.088$ & $0.818 \pm 0.130$ & $0.428 \pm 0.180$ & $0.607 \pm 0.103$ \\
        
        \cmidrule[0.2pt]{1-7}
        
        \multirow{5}{*}{\textbf{TwinMarket}} 
         & GPT-4o mini   & $0.025 \pm 0.031$ & $0.440 \pm 0.124$ & $0.510 \pm 0.290$ & $0.298 \pm 0.055$ & $0.591 \pm 0.067$ \\
         &  Llama-3.1-8B-Instruct  & $0.042 \pm 0.012$ & $0.439 \pm 0.047$ & $0.745 \pm 0.079$ & $0.325 \pm 0.029$ & $0.573 \pm 0.068$ \\
         & Claude-3-Haiku & $0.139 \pm 0.046$ & $0.460 \pm 0.074$ & $0.777 \pm 0.153$ & $0.350 \pm 0.039$ & $0.752 \pm 0.050$ \\
         & Qwen-Plus    & $0.216 \pm 0.268$ & $0.512 \pm 0.047$ & $0.813 \pm 0.107$ & $0.253 \pm 0.063$ & $0.655 \pm 0.041$ \\
         & DeepSeek-V3.2     & $0.047 \pm 0.042$ & $0.426 \pm 0.052$ & $0.645 \pm 0.109$ & $0.296 \pm 0.028$ & $0.657 \pm 0.064$ \\
        
        \cmidrule[0.2pt]{1-7}
        
        \multirow{5}{*}{\textbf{SocialNetwork}} 
         & GPT-4o    & $0.002 \pm 0.008$ & $0.360 \pm 0.042$ & $-0.129 \pm 0.280$ & $0.448 \pm 0.052$ & $0.654 \pm 0.034$ \\
         & Llama-3.1-8B-Instruct  & $0.000 \pm 0.000$ & $0.397 \pm 0.044$ & $0.276 \pm 0.370$ & $0.422 \pm 0.050$ & $0.647 \pm 0.047$ \\
         & Claude-3-Haiku & $0.011 \pm 0.016$ & $0.365 \pm 0.090$ & $0.280 \pm 0.516$ & $0.549 \pm 0.127$ & $0.616 \pm 0.139$ \\
         & Qwen-Plus   & $0.016 \pm 0.031$ & $0.391 \pm 0.066$ & $0.652 \pm 0.249$ & $0.727 \pm 0.108$ & $0.644 \pm 0.037$ \\
         & DeepSeek-V3.2    & $0.006 \pm 0.020$ & $0.334 \pm 0.096$ & $0.406 \pm 0.429$ & $0.420 \pm 0.154$ & $0.620 \pm 0.061$ \\

        \bottomrule
    \end{tabular}%
    }
    \vskip -10pt
\end{table*}

\subsection{When does the Extreme Event Originate?}\label{subsec:when}

\textbf{Evaluating the risk contribution $\phi_t^{\text{tm}}$ of each time step $t$ to the extreme event.}  Our goal is to identify the key time step by which the majority of the risk has already been accumulated. As shown in Figure~\ref{fig:metric}(a), we first aggregate the attribution scores of all actions $\phi(a_{i,t})$ at each time step $t$ to measure the temporal risk contribution as:
\begin{equation}
\phi^{\text{tm}}_{t} = \sum_{i=1}^{N} \phi(a_{i,t}).
\end{equation}

As defined in Section~\ref{subsec:pre}, the extreme event happens when the event risk exceeds a threshold {\small $R_{T}(\tau)=\sum_{t=1}^T \phi^{\text{tm}}_{t}>\rho$}. We define $T^*$ as the first time step where the accumulation of temporal contribution {\small$\sum_{t=1}^{t'} \phi^{\text{tm}}_{t}$} exceeds a certain proportion $q\in \mathbb{R}$ of the threshold, \emph{i.e.}, {\small$T^* = \min \{t' \mid \sum_{t=1}^{t'} \phi^{\text{tm}}_{t} > q\cdot\rho \}$}. In this paper, we set $q=0.9$ unless otherwise stated. While the extreme event finally happens at time $T$, the majority of its associated risk is mostly accumulated before time $T^*$. We further design a metric to quantitatively analyze the gap between $T^*$ and $T$.

\begin{definition} \itshape
    \textbf{(Metric: relative risk latency $L_{\text{tm}}$)} To evaluate the earliness of risk formation relative to the entire event duration, we define the relative risk latency $L_{\text{tm}}$ as the normalized difference between the time when the event occurs $T$ and the time when the risk is mostly formed $T^*$ as follows.
    \begin{equation}\label{eq:time}
        L_{\text{tm}} = (T - T^*)/T \in (0,1],
    \end{equation}
    where {\small $T^* = \min \{t' \mid \sum_{t=1}^{t'} \phi^{\text{tm}}_{t} > q \cdot \rho \}$} denotes the time step when the majority of risk is accumulated.
\end{definition}

\begin{figure*}[!t]
  \begin{center}
    \centerline{\includegraphics[width=\linewidth]{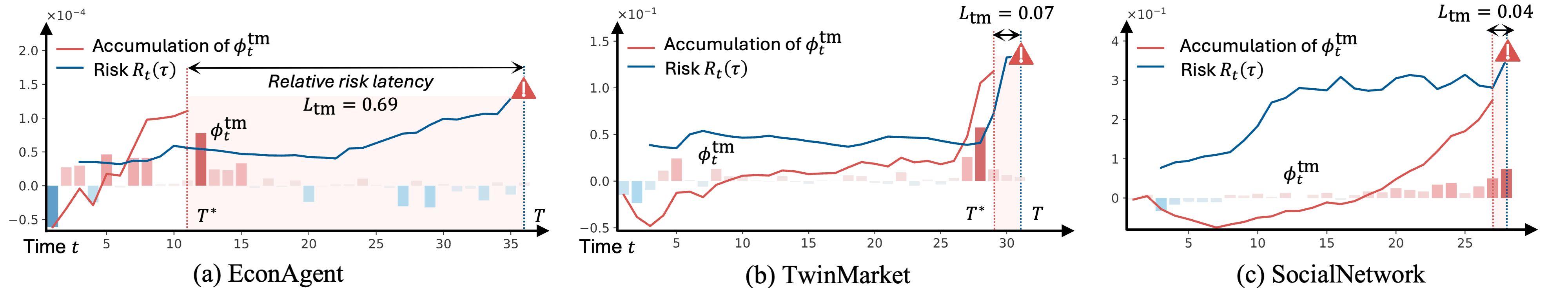}}
    \caption{Visualization of risk latency. The bar chart represents the risk contribution $\phi^{\text{tm}}_{t}$ of all actions taken at time $t$. The red line denotes the cumulative risk $\sum_{t'=1}^{t}\phi^{\text{tm}}_{t'}$, while the blue line represents the observed risk $R_{t}(\tau)$ in the real world at time $t$.}
    \label{fig:temporal}
  \end{center}
    \vskip -15pt
\end{figure*}

As Figure~\ref{fig:temporal} shows, this metric evaluates the temporal latency of the extreme event. A large $L_{\text{tm}}$ indicates that the risk originates early and undergoes a long dormant phase before the outbreak, whereas a small $L_{\text{tm}}$ implies a sudden shock where the risk forms immediately before the event.

\begin{ExampleBox}
\textbf{\textit{Insight 1. Extreme events originate with distinct temporal patterns: either from early dormant risks or immediate shocks.}} 
\end{ExampleBox} 

We followed the settings in Section~\ref{subsec:setup} to test the relative risk latency $L_{\text{tm}}$ in various MAS. As Table~\ref{tab:metric} shows, we found that EconAgent exhibits high latency (often $L_{\text{tm}} >0.6$), signaling a long dormant phase where risk stays hidden. Conversely, TwinMarket and SocialNetwork ($L_{\text{tm}} \approx0$) are shock-driven, with risks triggering immediate outbreaks.

\begin{figure*}[!t]
  \begin{center}
    \centerline{\includegraphics[width=\linewidth]{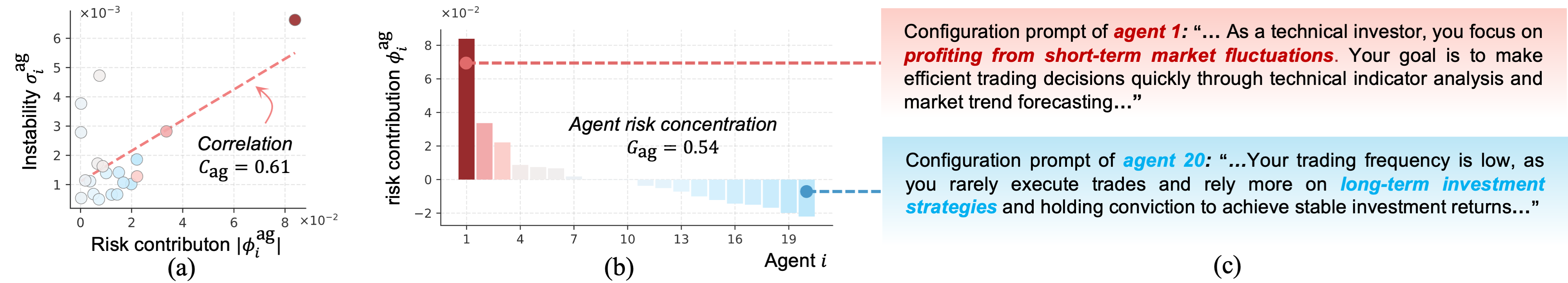}}
    \caption{Visualization of risk from each agent. The agents are sorted by the value of $\phi_i^{\text{ag}}$. (a) Correlation between agent instability and absolute risk contribution ($C_{\text{ag}} = 0.61$). (b) Distribution of risk contributions $\phi^{\text{ag}}_i$ across agents, showing significant risk concentration ($G_{\text{ag}} = 0.54$). (c) Configuration prompts for high-risk (agent 1, short-term) and low-risk (agent 20, long-term) investment strategies. }
    \label{fig:agent}
  \end{center}
  \vskip -20pt
\end{figure*} 

\textbf{Visualizing temporal risk contributions $\phi_t^{\text{tm}}$ in MAS.} As shown in Figure~\ref{fig:temporal}, we visualized the contribution $\phi^{\text{tm}}_{t}$ of all actions, the accumulated risk $\sum_{t'=1}^{t}\phi^{\text{tm}}_{t'}$, and the real risk $R_t(\tau)$ at time step $t$ to illustrate the risk accumulation process over time. Three trajectories were visualized using the GPT-4o mini model on scenarios of EconAgent, TwinMarket and SocialNetwork, respectively.

\subsection{Who Drives the Extreme Event?}\label{subsec:who}

\textbf{Evaluating the risk contribution $\phi^{\text{ag}}_i$ of each agent  $i$ to the extreme event.}  Our goal is to identify key agents based on their risk contribution. As shown in Figure~\ref{fig:metric}, we aggregate the attribution scores of all actions $\phi(a_{i,t})$ taken by agent $i$ to measure the total risk contribution of agent $i$. We also calculate the contribution standard deviation $\sigma^{\text{ag}}_i$ for each agent $i$ to measure its instability.
{\small \begin{equation}
    \phi^{\text{ag}}_i = \sum_{t=1}^{T} \phi(a_{i,t}), \quad \sigma^{\text{ag}}_i = \operatorname{std}\left( \phi(a_{i,1}), \dots, \phi(a_{i,T}) \right),
\end{equation}}
where $\operatorname{std}(\cdot)$ denotes the standard deviation. To further characterize the features of the risk contributions of agents, we design three metrics as follows. 

\begin{definition} \itshape
    \textbf{(Metric: agent risk concentration $G_{\text{ag}}$)} To evaluate the degree of risk concentration among agents, we define the agent risk concentration $G_{\text{ag}}$ based on the Gini Coefficient~\cite{hurley2009comparing} as follows:
    {\small \begin{equation}
        G_{\text{ag}}= \frac{\sum_{i=1}^{N}\sum_{j=1}^{N}||\phi^{\text{ag}}_i|-|\phi^{\text{ag}}_j||}{2N \sum_{i=1}^{N}|\phi^{\text{ag}}_i|} \in [0,1],
    \end{equation}}
    where $|\cdot|$ denotes taking the absolute value.
\end{definition}

$G_{\text{ag}}$ measures the concentration of risk contribution. A higher $G_{\text{ag}}$ indicates that the risk is driven by a small minority of agents, while a lower $G_{\text{ag}}$ implies that the risk is distributed evenly among the agents.

\begin{ExampleBox}
\textbf{\textit{Insight 2. Extreme events are typically driven by a small subset of agents.}} 
\end{ExampleBox}
We evaluated the agent risk concentration $G_{\text{ag}}$ across various LLMs and three scenarios following the setup in Section~\ref{subsec:setup}. Table~\ref{tab:metric} reveals that the concentration levels are consistently high (often $G_{\text{ag}}>0.4$). This confirms that risk is not a system-wide average, but is predominantly generated by a handful of critical agents.

\begin{definition} \itshape
    \textbf{(Metric: risk-instability correlation $C_{\text{ag}}$)} To quantify the correlation between an individual agent's risk contribution and its instability, we define the risk-instability correlation as the Pearson correlation coefficient~\cite{lee1988thirteen} between the risk contribution $\phi^{\text{ag}}_i$ and the instability $\sigma^{\text{ag}}_i$:
    {\small\begin{equation}
        C_{\text{ag}} = \frac{\operatorname{Cov}(\boldsymbol{\phi}, \boldsymbol{\sigma})}{\operatorname{std}(\boldsymbol{\phi}) \cdot \operatorname{std}(\boldsymbol{\sigma})} \in [-1,1],
    \end{equation}}
    where $\boldsymbol{\phi} = [|\phi^{\text{ag}}_1|, \dots, |\phi^{\text{ag}}_N|]$ and $\boldsymbol{\sigma} = [\sigma^{\text{ag}}_1, \dots, \sigma^{\text{ag}}_N]$ are the vectors of risk contribution and instability across all agents. Here, $\operatorname{Cov}(\cdot)$ denotes the covariance, $\operatorname{std}(\cdot)$ denotes the standard deviation, and $|\cdot|$ denotes the absolute value.
\end{definition}

A large positive $C_{\text{ag}}$ indicates that the most dangerous agents are also the most unstable ones as shown in Figure~\ref{fig:agent}(a). In contrast, a large negative $C_{\text{ag}}$ indicates that the highest risks are contributed by agents who act stably over time.

\begin{ExampleBox}
\textbf{\textit{Insight 3. Agents with high risk contribution often exhibit high instability.}} 
\end{ExampleBox}
We tested the risk-instability correlation $C_{\text{ag}}$ following the setup in Section~\ref{subsec:setup}. Table~\ref{tab:metric} shows consistently high positive values (often $C_{\text{ag}}>0.6$). This strong correlation confirms that the most dangerous agents are not steady actors, but rather those exhibiting high instability.

\begin{definition} \itshape
    \textbf{(Metric: agent risk synchronization $Z_{\text{ag}}$)} To quantify the synchronization of agent risk contributions, we define the agent risk synchronization $Z_{\text{ag}}$ by adapting the order parameter from the Vicsek model~\cite{vicsek1995novel}:
    {\small\begin{equation}
        Z_{\text{ag}} = \mathbb{E}_{t} \big[\ \frac{|\sum_{i=1}^{N}\phi(a_{i,t})|}{\sum_{i=1}^{N}|\phi(a_{i,t})|}\ \big] \in [0,1],
    \end{equation}}
    where $\mathbb{E}_{t}[\cdot]$ denotes taking the average over all time steps and  $|\cdot|$ denotes taking the absolute value.
\end{definition}

\begin{figure*}[!t]
  \begin{center}
    \centerline{\includegraphics[width=\linewidth]{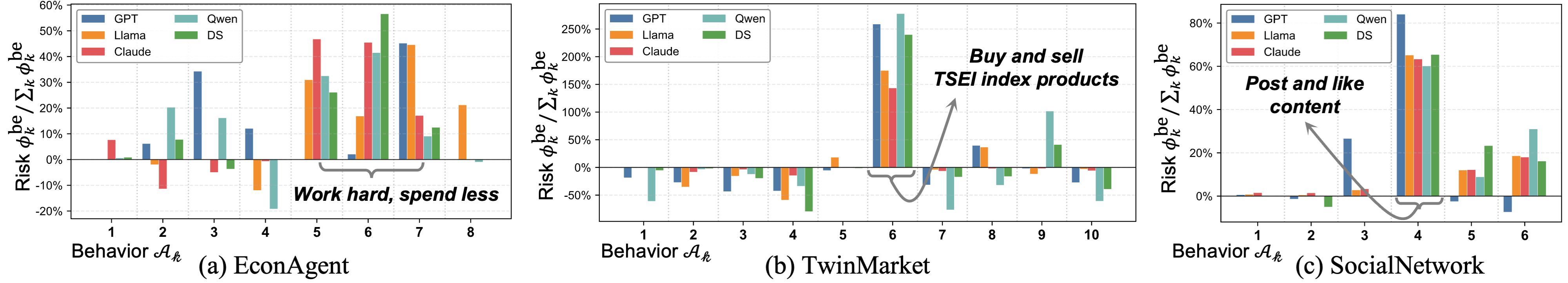}}
    \caption{Visualization of the risk contribution $\phi^{\text{be}}_{k}$ of different behaviors $\mathcal{A}_k$. The contributions are normalized such that the total equals 100\%. We also highlight the high-risk behaviors in each system. GPT, Llama, Claude, Qwen, and DS denote the results of model GPT-4o mini, Llama-3.1-8B-Instruct, Claude-3-Haiku, Qwen-Plus, and DeepSeek-V3.2, respectively.}
    \label{fig:action}
  \end{center}
    \vskip -20pt
\end{figure*}

The metric $Z_{\text{ag}}$ measures the directional alignment of risk contributions. A high $Z_{\text{ag}}$ indicates that agents contribute to risk synchronously (all increasing or decreasing together), whereas a low $Z_{\text{ag}}$ implies that positive and negative contributions cancel each other out.

\begin{ExampleBox}
\textbf{\textit{Insight 4. Agents tend to increase or decrease risk synchronously.}} 
\end{ExampleBox}
We evaluated the risk synchronization $Z_{\text{ag}}$ following the setup in Section~\ref{subsec:setup}. As Table~\ref{tab:metric} shows, the synchronization scores are generally high (often $Z_{\text{ag}}>0.3$). This indicates that agents tend to move in the same direction, \emph{i.e.}, increasing or decreasing risk synchronously.

\textbf{Visualizing agent risk contributions $\phi_i^{\text{ag}}$ in MAS.} As shown in Figure~\ref{fig:agent}(b), given an extreme event case using the Qwen-Plus model on the TwinMarket, we visualized the risk contribution $\phi^{\text{ag}}_{i}$ of all actions performed by agent $i$.  We also visualized the correlation between the absolute value of the contribution $|\phi^{\text{ag}}_{i}|$ and instability $\sigma^{\text{ag}}_{i}$, as shown in Figure~\ref{fig:agent}(a). Finally, we visualized the configuration prompts for the high-risk agent $i=1$ and the low-risk agent $i=20$ with different investment strategies in Figure~\ref{fig:agent}(c).

\subsection{What Behaviors Contribute to the Extreme Event?}\label{subsec:what}

\textbf{Evaluating the risk contribution $\phi^{\text{be}}_k$ of each action pattern $\mathcal{A}_k$ to the extreme event.} Let $\mathcal{A}_k$ denote the $k$-th type of action pattern\footnote{The types of behaviors are defined by domain experts. Please refer to Appendix~\ref{app:scenario} for the definition for the specific system.}. As shown in Figure~\ref{fig:metric}, we calculate the behavioral risk contribution $\phi^{\text{be}}_k$ as the sum of attribution scores from all instances where this action pattern occurs as follows:
\begin{equation}
    \phi^{\text{be}}_k = \sum_{t=1}^{T} \sum_{i=1}^{N} \mathbb{I}(a_{i,t} \in \mathcal{A}_{k}) \cdot \phi(a_{i,t}),
\end{equation}
where $\mathbb{I}(\cdot)$ is the indicator function, which equals $1$ if $a_{i,t} \in \mathcal{A}_{k}$ and $0$ otherwise. To further characterize the behavioral feature of the extreme event, we introduce the following metric.

\begin{definition} \itshape
   \textbf{(Metric: behavior risk concentration $G_{\text{be}}$)} To quantify the concentration of risk contributions among different action patterns, we define the behavior risk concentration $G_{\text{be}}$ using the Gini Coefficient~\cite{hurley2009comparing} of the behavioral risk contributions:
    {\small\begin{equation}
        G_{\text{be}}= \frac{\sum_{s=1}^{K}\sum_{k=1}^{K}||\phi^{\text{be}}_s|-|\phi^{\text{be}}_k||}{2K \sum_{k=1}^{K}|\phi^{\text{be}}_k|} \in [0,1],
    \end{equation}}
    where $K$ denotes the number of the total types of behaviors and $|\cdot|$ denotes taking the absolute value.
\end{definition}

A high $G_{\text{be}}$ indicates that the systemic risk is concentrated in a few critical behaviors, whereas a low $G_{\text{be}}$ suggests that the risk arises from a combination of various action patterns. 

\begin{ExampleBox}
\textbf{\textit{Insight 5. A small number of behaviors contribute the majority of the risk leading to extreme events.}} 
\end{ExampleBox}
We evaluated the behavior concentration $G_{\text{be}}$ following the setup in Section~\ref{subsec:setup}. As Table~\ref{tab:metric} shows, the behavior concentration is consistently high (often $G_{\text{be}}>0.5$). This confirms that risk is not randomly distributed across all actions, but stems from a few dominant, stereotypical behaviors.

\textbf{Visualizing behavioral risk contributions $\phi^{\text{be}}_k$ in MAS.} As shown in Figure~\ref{fig:action}, we visualized the contribution $\phi^{\text{be}}_{k}$ of each action pattern, with a specific focus on the high-risk action patterns in each scenario. We conduct experiments using five representative LLMs across three scenarios of EconAgent, TwinMarket and SocialNetwork. For each experimental setting, the results were averaged over five independent trajectories to ensure reliability. Please refer to Section~\ref{subsec:setup} for detailed settings and see Appendix~\ref{app:scenario} for details on behavior definitions for each scenario.

\begin{table*}[!t]
    \centering
    \caption{Faithfulness evaluation of attribution methods in MAS. We report the risk drop $(R_T(\tau)-R_T(\tau^{del}))/R_T(\tau)$ in percentage (\%) after deleting the top-3 and top-10 most dangerous actions $a_{i,t}$ (those with the highest attribution scores $\phi(a_{i,t})$). Higher values indicate more faithful attribution.}
    \label{tab:faithfulness}
    \resizebox{\linewidth}{!}{%
    \begin{tabular}{ll cc cc cc cc cc}
        \toprule
        \multirow{2}{*}{\textbf{Scenario}} & \multirow{2}{*}{\textbf{Method}} & \multicolumn{2}{c}{\textbf{GPT-4o mini}} & \multicolumn{2}{c}{\textbf{Llama-3.1-8B}} & \multicolumn{2}{c}{\textbf{Claude-3-Haiku}} & \multicolumn{2}{c}{\textbf{Qwen-Plus}} & \multicolumn{2}{c}{\textbf{DeepSeek-V3.2}} \\
        \cmidrule(lr){3-4} \cmidrule(lr){5-6} \cmidrule(lr){7-8} \cmidrule(lr){9-10} \cmidrule(lr){11-12}
        & & Top-3 & Top-10 & Top-3 & Top-10 & Top-3 & Top-10 & Top-3 & Top-10 & Top-3 & Top-10 \\
        \midrule
        \multirow{5}{*}{\textbf{EconAgent}} 
        & Random                       & -3.50 & 7.69  & -13.75 & 0.64  & 1.73  & 20.39 & -86.35 & -16.53 & -5.76 & 10.53 \\
        & FT~\cite{cemri2025multi}     & -18.25 & -19.05 & -7.39  & -3.06 & 25.44 & 41.24 & -55.09 & -33.09 & -24.82 & -9.94 \\
        & FA~\cite{zhang2025agent}     & 1.49  & 14.14 & -13.99 & 7.73  & 25.92 & 40.37 & -70.63 & -53.92 & -21.57 & -3.23 \\
        & AT~\cite{zhang2025agentracer} & 5.32  & 27.49 & 20.18 & 31.89 & 34.95 & 39.17 & -69.71 & -27.61 & -18.20 & -0.59 \\
        & \textbf{Ours}                & \textbf{36.31} & \textbf{45.42} & \textbf{36.28} & \textbf{47.44} & \textbf{45.72} & \textbf{57.08} & \textbf{-32.09} & \textbf{1.32} & \textbf{11.56} & \textbf{32.53} \\
        \midrule
        \multirow{5}{*}{\textbf{TwinMarket}} 
        & Random                       & 0.49  & 1.39  & 1.15  & -4.94 & 0.49  & 2.77  & 0.43  & 1.00  & -0.27 & 8.36  \\
        & FT~\cite{cemri2025multi}     & -1.34 & 4.64  & -1.17 & -3.38 & -1.34 & 0.57  & 0.79  & 0.62  & -0.14 & 2.03  \\
        & FA~\cite{zhang2025agent}     & -1.43 & 4.76  & 0.27  & -0.39 & -1.43 & 0.55  & -0.17 & -0.51 & -0.14 & -8.91 \\
        & AT~\cite{zhang2025agentracer} & 16.63 & 21.76 & \textbf{44.91} & 41.12 & 16.63 & 63.56 & \textbf{65.14} & 46.06 & 25.50 & 40.72 \\
        & \textbf{Ours}                & \textbf{38.20} & \textbf{60.05} & 17.00 & \textbf{50.88} & \textbf{38.20} & \textbf{77.90} & 52.82 & \textbf{55.54} & \textbf{46.21} & \textbf{56.46} \\
        \midrule
        \multirow{5}{*}{\textbf{SocialNetwork}} 
        & Random                       & 12.2 & 15.1 & 8.0 & 11.4 & 2.9 & 2.7 & 2.8 & 2.9 & 1.9 & 2.6 \\
        & FT~\cite{cemri2025multi}     & 12.9 & 14.9 & 8.3 & 10.8 & 2.3 & 2.1 & 2.7 & 3.0 & -0.1 & 1.5 \\
        & FA~\cite{zhang2025agent}     & 13.0 & 16.6 & 8.3 & 11.1 & 2.8 & 2.2 & 2.7 & 3.5 & -0.1 & 3.1 \\
        & AT~\cite{zhang2025agentracer} & \textbf{22.2} & 39.3 & \textbf{19.0} & \textbf{32.7} & 11.1 & 19.2 & \textbf{7.4} & 8.5 & 5.9 & 10.6 \\
        & \textbf{Ours}                & 17.9 & \textbf{39.7} & 11.9 & 25.5 & \textbf{11.9} & \textbf{23.1} & 4.6 & \textbf{8.9} & \textbf{10.4} & \textbf{11.6} \\
        \bottomrule
    \end{tabular}
    }
    \vskip -10pt
\end{table*}

\section{Experiments}

\subsection{Set Up}\label{subsec:setup}
For each experimental setting, we executed the MAS using different random seeds to generate five independent trajectories where extreme events occurred. All reported experimental results were derived from these five trajectories unless otherwise stated. 

\textbf{Models.} We conducted experiments using representative LLMs, including GPT-4o mini~\cite{hurst2024gpt}, Llama-3.1-8B-Instruct~\cite{grattafiori2024llama}, Claude-3-Haiku~\cite{anthropic2024claude}, Qwen-Plus~\cite{yang2025qwen3}, and DeepSeek-V3.2~\cite{liu2024deepseek}. All LLM APIs were accessed using default parameters.

\textbf{MAS scenarios.} We evaluated three LLM-powered MAS scenarios covering economic systems, financial markets, and social networks. The first scenario is \textbf{EconAgent}~\cite{li2024econagent}, which simulates a macroeconomic system. Each agent takes action $a_{i,t}$ to decide whether to work or consume at each time step. We adopt the RiskMetrics standard~\cite{BOLLERSLEV1986307} to measure economic uncertainty risk $R_t(\tau)$, by computing the conditional variance of inflation forecast errors. For this scenario, we set the number of agents to $N = 10$, with an average trajectory length of $\overline{T}=33.8$ (please see Appendix~\ref{app:econagent} for details). The second scenario is \textbf{TwinMarket}~\cite{yang2025twinmarket}, which simulates a financial market. Agents take actions $a_{i,t}$ to buy, sell, or hold a quantity of the index. Similarly, we adopt the RiskMetrics standard~\cite{BOLLERSLEV1986307} to measure market risk $R_t(\tau)$, by computing the conditional variance of forecast errors for composite market index returns. For this scenario, we set $N = 10$, with an average trajectory length of $\overline{T}=27.2$ (see Appendix~\ref{app:twinmarket} for details). The third scenario is \textbf{SocialNetwork}~\cite{stauffer2004simulation}, which simulates a social platform. Agents take actions $a_{i,t}$ to like, dislike, or not interact with posted information. The risk $R_t(\tau)$ is defined as the variance of agent beliefs~\cite{BOLLERSLEV1986307,bramson2017understanding}, quantifying the polarization level of the network. For this scenario, we set $N = 20$, with an average trajectory length of {$\overline{T}=20.6$} (see Appendix~\ref{app:socialnetwork} for details).

\subsection{Evaluating the Faithfulness of the Attribution}\label{subsec:faithfulness}

\textbf{Metrics.} We conducted experiments to compare the faithfulness of the proposed attribution method with competing methods designed for MAS. Given the trajectory $\tau$ leading to the extreme event and the corresponding attribution scores $\phi(a_{i,t})$ for each action $a_{i,t}$, we selected the top-3 and the top-10 actions with the highest attribution scores to represent the identified dangerous actions. These dangerous actions were then deleted to generate a new trajectory $\tau^{\text{del}}$, following the procedure in Section~\ref{subsec:pre}. We used the final risk drop, defined as $(R_{T}(\tau)-R_{T}(\tau^{\text{del}}))/R_{T}(\tau)$, to measure the faithfulness of the attribution method.  The other experimental settings are described in Section~\ref{subsec:setup}.

\textbf{Baselines.} We compared our method with four competing baselines. The first is Random, where we generated scores randomly from a normal distribution as a basic baseline. The second and third methods are Failure Taxonomy (FT)~\cite{cemri2025multi} and Failure Attribution (FA)~\cite{zhang2025agent}, which prompt an LLM to get attribution scores (please refer to Appendix~\ref{app:ft} and Appendix~\ref{app:fa} for technical details). The fourth method is Agent Tracer (AT)~\cite{zhang2025agentracer}, a counterfactual-based method for estimating agent contribution (please refer to Appendix~\ref{app:at} for technical details). In comparison, we adapted the Shapley value to compute the attribution scores as introduced in Section~\ref{subsec:pre}.

\textbf{Results.} Table~\ref{tab:faithfulness} shows the experimental results. We found  our method achieved the highest risk drop across the majority of experimental settings. The results demonstrate that, compared to other competing methods designed for MAS, the method based on the Shapley value faithfully reflects the influence of the actions.

\section{Conclusion}

\label{sec:conclusion}

In this paper, we proposed a framework for explaining emergent extreme events in LLM-powered MAS. Based on the Shapley value, this approach allows for the faithful quantification of the contribution of each agent's action to the extreme event. By aggregating these  attribution scores, we constructed a multi-dimensional analysis framework to interpret extreme events from the perspectives of \textit{when}, \textit{who}, and \textit{what}.

Since the computation of the Shapley value is NP-hard, we applied a Monte Carlo approximation to effectively handle scenarios involving thousands of actions. Scaling the framework to significantly larger systems remains a primary objective for our future work. Nevertheless, experiments across diverse MAS scenarios demonstrate the effectiveness of our framework and provide general insights into the emergence of extreme phenomena.

\section*{Impact Statement}

This paper presents work whose goal is to advance the field of Machine
Learning. There are many potential societal consequences of our work, none
which we feel must be specifically highlighted here.

\bibliography{example_paper}
\bibliographystyle{icml2026}

\newpage
\appendix
\onecolumn

\section{Thoeretical analisis}
\subsection{Properties of the Shapley Value for Extreme Event Attribution}
\label{app:shapley}

In this subsection, we demonstrate that the Shapley value used in our framework satisfies the four axiomatic properties: \textit{Efficiency}, \textit{Symmetry}, \textit{Nullity}, and \textit{Linearity}. These properties theoretically guarantee that the attribution of the final risk $R_T(\tau)$ to individual actions $a_{i,t} \in \Omega$ is unique and fair.

To align the Shapley value axioms with our risk decomposition, we define the characteristic function $v(S)$ as the \textbf{risk deviation} of a subset of actions $S$ relative to the safe baseline:
\begin{equation}
    v(S) = R_T(\tau^S) = \hat{R}_T(\tau^S) - \hat{R}_T(\tau^\emptyset),
\end{equation}
where $\hat{R}_T(\tau^S)$ is the raw risk metric evaluated on the counterfactual trajectory $\tau^S$, and $\tau^\emptyset$ represents the baseline trajectory where all actions are replaced by expert-annotated safe actions. By this definition, it naturally follows that \textbf{$v(\emptyset) = 0$}. The total risk to be decomposed in our framework is thus the total deviation $R_T(\tau) = v(\Omega) = \hat{R}_T(\tau^\Omega) - \hat{R}_T(\tau^\emptyset)$.

Recall the definition of the Shapley value for an action $a_{i,t}$ given in Eq.~\eqref{eq:shapley}:
\begin{equation}
    \phi_v(a_{i,t}) = \sum_{S \subseteq \Omega \setminus \{a_{i,t}\}} \frac{|S|!(|\Omega|-|S|-1)!}{|\Omega|!} \left( v(S \cup \{a_{i,t}\}) - v(S) \right).
\end{equation}

\vspace{5pt}
\noindent \textbf{Property 1 (Efficiency).} The total attribution sums up to the total risk.
\begin{equation}
    \sum_{a_{i,t} \in \Omega} \phi_v(a_{i,t}) = R_T(\tau).
\end{equation}
\begin{proof}
Let $\Pi(\Omega)$ be the set of all possible permutations of the action set $\Omega$, where each permutation $\pi \in \Pi(\Omega)$ represents a specific ordering of adding actions to the game. The Shapley value can be equivalently interpreted as the average marginal contribution over all permutations:
\begin{equation}
    \phi_v(a_{i,t}) = \frac{1}{|\Omega|!} \sum_{\pi \in \Pi(\Omega)} \left( v(Pre^i(\pi) \cup \{a_{i,t}\}) - v(Pre^i(\pi)) \right),
\end{equation}
where $Pre^i(\pi)$ denotes the set of actions preceding $a_{i,t}$ in permutation $\pi$. Summing over all actions:
\begin{equation}
    \sum_{a_{i,t} \in \Omega} \phi_v(a_{i,t}) = \frac{1}{|\Omega|!} \sum_{\pi \in \Pi(\Omega)} \sum_{a_{i,t} \in \Omega} \left( v(Pre^i(\pi) \cup \{a_{i,t}\}) - v(Pre^i(\pi)) \right).
\end{equation}
For any specific permutation $\pi$, the inner sum forms a telescoping series that starts at $v(\emptyset)$ and ends at $v(\Omega)$. Given our construction $v(\emptyset) = 0$, the equation simplifies to:
\begin{equation}
    \sum_{a_{i,t} \in \Omega} \phi_v(a_{i,t}) = \frac{1}{|\Omega|!} \sum_{\pi \in \Pi(\Omega)} (v(\Omega) - 0) = v(\Omega) = R_T(\tau).
\end{equation}
This ensures that the sum of attributions perfectly explains the extreme event risk relative to the safe baseline.
\end{proof}

\vspace{5pt}
\noindent \textbf{Property 2 (Symmetry).} If two actions $a_{i,t}$ and $a_{j,k}$ make identical contributions to any subset of actions, they receive equal attribution.
\begin{equation}
    \forall S \subseteq \Omega \setminus \{a_{i,t}, a_{j,k}\}, \ v(S \cup \{a_{i,t}\}) = v(S \cup \{a_{j,k}\}) \implies \phi_v(a_{i,t}) = \phi_v(a_{j,k}).
\end{equation}
\begin{proof}
Let $w_{|S|} = \frac{|S|!(|\Omega|-|S|-1)!}{|\Omega|!}$. The Shapley value depends only on the weight $w_{|S|}$ and the marginal contribution $v(S \cup \{a_{i,t}\}) - v(S)$. Since $w_{|S|}$ is determined solely by the cardinality of $S$, and by hypothesis the marginal contributions are identical for all $S \subseteq \Omega \setminus \{a_{i,t}, a_{j,k}\}$, it follows directly from Eq.~\eqref{eq:shapley} that $\phi_v(a_{i,t}) = \phi_v(a_{j,k})$.
\end{proof}

\vspace{5pt}
\noindent \textbf{Property 3 (Nullity).} If an action $a_{i,t}$ has no impact on the risk in any context, its attribution is zero.
\begin{equation}
    \forall S \subseteq \Omega \setminus \{a_{i,t}\}, \ v(S \cup \{a_{i,t}\}) = v(S) \implies \phi_v(a_{i,t}) = 0.
\end{equation}
\begin{proof}
Given the condition $v(S \cup \{a_{i,t}\}) = v(S)$, the marginal contribution term in Eq.~\eqref{eq:shapley} becomes $v(S \cup \{a_{i,t}\}) - v(S) = 0$. Substituting this into the Shapley summation:
\begin{equation}
    \phi_v(a_{i,t}) = \sum_{S \subseteq \Omega \setminus \{a_{i,t}\}} w_{|S|} \cdot 0 = 0.
\end{equation}
In our context, this confirms that actions that do not increase the risk beyond the baseline level receive zero attribution.
\end{proof}

\vspace{5pt}
\noindent \textbf{Property 4 (Linearity).} The Shapley value of a linear combination of games is the linear combination of the Shapley values.
\begin{equation}
    \text{Let } v = \alpha v_1 + \beta v_2, \text{ then } \phi_v(a_{i,t}) = \alpha \phi_{v_1}(a_{i,t}) + \beta \phi_{v_2}(a_{i,t}).
\end{equation}
\begin{proof}
Substituting the composite characteristic function $v(S) = \alpha v_1(S) + \beta v_2(S)$ into Eq.~\eqref{eq:shapley}:
\begin{align}
    \phi_v(a_{i,t}) &= \sum_{S} w_{|S|} \left[ (\alpha v_1(S \cup \{a_{i,t}\}) + \beta v_2(S \cup \{a_{i,t}\})) - (\alpha v_1(S) + \beta v_2(S)) \right] \\
    &= \sum_{S} w_{|S|} \left[ \alpha (v_1(S \cup \{a_{i,t}\}) - v_1(S)) + \beta (v_2(S \cup \{a_{i,t}\}) - v_2(S)) \right] \\
    &= \alpha \phi_{v_1}(a_{i,t}) + \beta \phi_{v_2}(a_{i,t}).
\end{align}
This property allows us to decompose complex risk metrics and attribute them separately.
\end{proof}

\subsection{Estimating Shapley Values via Monte Carlo Sampling}
\label{app:mc}

In this subsection, we detail the Monte Carlo sampling algorithm used to approximate the Shapley values. Computing the exact Shapley value (Eq.~\eqref{eq:shapley}) requires summing over $2^{|\Omega|}$ subsets, which is computationally intractable for long trajectories where $|\Omega| = NT$ is large. To address this, we utilize a permutation-based approximation method that reduces the complexity from exponential to linear with respect to the number of actions.

\subsubsection{Permutation-based Formulation}
The Shapley value can be equivalently formulated as the expected marginal contribution of an action over all possible execution orders (permutations). Let $\Pi(\Omega)$ denote the set of all possible permutations of $\Omega$. For a given permutation $\pi \in \Pi(\Omega)$, let $Pre^i(\pi)$ be the set of actions that precede action $a_{i,t}$ in the ordering $\pi$. The Shapley value is given by:
\begin{equation}
    \phi(a_{i,t}) = \mathbb{E}_{\pi \sim U(\Pi(\Omega))} \left[ v(Pre^i(\pi) \cup \{a_{i,t}\}) - v(Pre^i(\pi)) \right],
\end{equation}
where $U(\Pi(\Omega))$ denotes the uniform distribution over all permutations. This expectation interpretation allows us to employ Monte Carlo sampling.

\subsubsection{Monte Carlo Approximation Algorithm}
Instead of iterating over all $|\Omega|!$ permutations, we approximate the expectation by sampling $M$ random permutations $\pi_1, \pi_2, \dots, \pi_M$. The approximate Shapley value $\hat{\phi}(a_{i,t})$ is calculated as the sample mean of the marginal contributions:
\begin{equation}
    \hat{\phi}(a_{i,t}) \approx \frac{1}{M} \sum_{m=1}^{M} \left( v(Pre^i(\pi_m) \cup \{a_{i,t}\}) - v(Pre^i(\pi_m)) \right).
\end{equation}

The detailed procedure is described in Algorithm~\ref{alg:shapley_mc}.

\begin{algorithm}[H]
\caption{Monte Carlo Estimation of Shapley Values for Extreme Event Attribution}
\label{alg:shapley_mc}
\begin{algorithmic}[1]
\REQUIRE Set of actions $\Omega$, Number of samples $M$, Risk function $v(\cdot)$
\ENSURE Approximate attribution scores $\hat{\phi}(a_{i,t})$ for all $a_{i,t} \in \Omega$

\STATE Initialize $\hat{\phi}(a) \leftarrow 0$ for all $a \in \Omega$
\FOR{$m = 1$ to $M$}
    \STATE Sample a random permutation $\pi_m$ of $\Omega$
    \STATE $S \leftarrow \emptyset$ \COMMENT{Start with the baseline safe trajectory}
    \STATE $v_{prev} \leftarrow v(\emptyset)$
    \FOR{$j = 1$ to $|\Omega|$}
        \STATE Let $a_{curr} = \pi_m[j]$ be the $j$-th action in the permutation
        \STATE $S \leftarrow S \cup \{a_{curr}\}$ \COMMENT{Add action to the set}
        \STATE $v_{curr} \leftarrow v(S)$ \COMMENT{Evaluate risk via re-simulation}
        \STATE $marginal \leftarrow v_{curr} - v_{prev}$
        \STATE $\hat{\phi}(a_{curr}) \leftarrow \hat{\phi}(a_{curr}) + marginal$
        \STATE $v_{prev} \leftarrow v_{curr}$
    \ENDFOR
\ENDFOR
\STATE \textbf{return} $\hat{\phi}(a) / M$ for all $a \in \Omega$
\end{algorithmic}
\end{algorithm}

\subsubsection{Complexity Analysis}
\textbf{Proposition.} The time complexity of Algorithm~\ref{alg:shapley_mc} is $O(M \cdot |\Omega| \cdot C_{sim})$, where $C_{sim}$ is the cost of a single simulation evaluation.

\begin{proof}
The algorithm performs an outer loop of $M$ iterations. Inside each iteration, it traverses the permutation of length $|\Omega|$. For each step in the inner loop, one marginal contribution is calculated, requiring one evaluation of the characteristic function $v(S)$. Thus, the total number of evaluations is $M \times |\Omega|$. Compared to the exact calculation complexity of $O(2^{|\Omega|} \cdot C_{sim})$, the Monte Carlo approach is linear in terms of the number of players $|\Omega|$, making it feasible for high-dimensional action spaces typical in MAS.
\end{proof}

\section{Simulation Scenarios}\label{app:scenario}

\subsection{Macroeconomic Simulation: EconAgent}\label{app:econagent}
The \textbf{EconAgent} environment simulates a macroeconomic ecosystem based on a one-step economy scenario. In this system, multiple agents engage in economic activities through labor supply and consumption decisions, while the system dynamically computes aggregate indicators such as price levels, wages, and inflation rates. Each agent's action space comprises two primary dimensions: \textit{SimpleLabor}, where agents make a binary choice to supply labor (work) or not, and \textit{SimpleConsumption}, where agents select a specific consumption rate.

Each agent's action space consists of two dimensions: \textit{SimpleLabor}, a binary decision indicating whether the agent supplies labor, and \textit{SimpleConsumption}, representing the chosen consumption ratio. To analyze behavior risk concentration, agent behaviors are categorized into eight discrete action types, defined by the Cartesian product of labor participation (work or not work) and four consumption intervals: $[0,25\%)$, $[25,50\%)$, $[50,75\%)$, and $[75,100\%)$.

In Figure~\ref{fig:action}(a), actions are indexed according to this combination of labor participation and consumption intensity. Specifically, actions 1--4 correspond to agents choosing not to work, with consumption ratios falling into the four increasing intervals listed above, while actions 5--8 correspond to agents choosing to work, with the same consumption intervals ordered from low to high.

In counterfactual simulations, actions are not removed entirely but replaced by a predefined \emph{baseline action}, which represents a safe and economically neutral behavior. In the EconAgent environment, the baseline action is defined as supplying labor with probability one and consuming at a fixed proportion of $0.8$. This baseline is designed to reflect a stable macroeconomic participation pattern that avoids introducing artificial shocks while preserving basic economic activity. All counterfactual trajectories used for Shapley value computation substitute the selected agent-time action with this baseline behavior.

System risk is quantified using the conditional variance of inflation forecast errors based on a naive forecast approach. Let $\pi_t$ denote the system's price inflation rate at time $t$. The expected inflation rate is defined as $E_{t-1}[\pi_t] = \pi_{t-1}$, yielding a forecast error of $e_t = \pi_t - E_{t-1}[\pi_t]$. The risk indicator $\hat{R}_t(\tau)$ is computed using an Exponential Weighted Moving Average (EWMA) model:
\begin{equation}
    \hat{R}_t(\tau) = \lambda \cdot \hat{R}_{t-1}(\tau) + (1-\lambda) \cdot e_{t-1}^2
\end{equation}
where $\lambda = 0.94$, consistent with the RiskMetrics standard. This metric captures the volatility of inflation dynamics and serves as a proxy for systemic macroeconomic risk.

Extreme events in the EconAgent environment are defined based on this risk indicator. Rather than relying on arbitrary statistical cutoffs, the threshold for extreme event realization is determined through expert annotation, incorporating domain knowledge regarding economically meaningful instability. 

\subsection{Financial Market Simulation: TwinMarket}\label{app:twinmarket}
\textbf{TwinMarket} simulates a financial trading system in which agents manage portfolios across ten distinct industry-level stock indices. At each timestep, an agent makes trading decisions independently for each index. For a given index, the agent may choose to \textit{buy} (by specifying order quantity and limit price), \textit{sell} (liquidating its current holdings), or \textit{hold} (taking no action).

For analysis and attribution purposes, agent behaviors are categorized by trading direction and asset type. The ten tradable indices, ordered consistently throughout the experiments, in Figure~\ref{fig:action}(b) are:
\texttt{TLEI}, \texttt{MEI}, \texttt{CPEI}, \texttt{IEEI}, \texttt{REEI}, \texttt{TSEI}, \texttt{CGEI}, \texttt{TTEI}, \texttt{EREI}, and \texttt{FSEI}.
Accordingly, actions indexed from 1 to 10 correspond to trading decisions on these indices in the listed order. For each indexed action, the agent may execute a buy or sell operation on the corresponding index, or choose to hold.

To enable counterfactual attribution, actions are not removed but replaced by a predefined baseline behavior. In the TwinMarket environment, the baseline action is defined as \textit{hold}, meaning that the agent maintains its current portfolio positions and does not generate any trading orders. This baseline represents a neutral and low-risk trading behavior and is used consistently across all counterfactual simulations for Shapley value computation.

The risk modeling framework in TwinMarket mirrors that of EconAgent, employing an EWMA-based volatility measure. System risk is derived from the log returns of a composite market price index $P_t$, which is constructed by aggregating the ten sector indices using VWAP or close-price weighting. The price change rate is defined as the log return:
\begin{equation}
    \pi_t = \ln P_t - \ln P_{t-1}.
\end{equation}
Using a naive forecast $E_{t-1}[\pi_t] = \pi_{t-1}$, the forecast error is computed and the risk indicator $\hat{R}_t(\tau)$ is obtained in the same manner as Eq.~(1), with decay factor $\lambda = 0.94$.

Extreme events are defined based on this risk indicator. The threshold is determined through expert annotation and specifies a risk level above which the market is considered to have entered an extreme or unstable state.

\subsection{Social Dynamics Simulation: SocialNetwork}\label{app:socialnetwork}
\textbf{SocialNetwork} models the dynamics of opinion polarization within a social media network, where agents interact through content creation and engagement. Each agent's action space consists of two main dimensions: \textit{posting}, a binary decision indicating whether the agent publishes content, and \textit{interaction}, where agents respond to each viewed post by selecting one of three options: \textit{like}, \textit{dislike}, or \textit{not interact}. Consequently, for each viewed post, the agent has six possible action types corresponding to the combination of posting and interaction decisions.

For behavior analysis, agent actions are categorized into six discrete types corresponding to the Cartesian product of posting and interaction decisions: \textit{post--like}, \textit{post--not interact}, \textit{post--dislike}, \textit{not post--like}, \textit{not post--not interact}, and \textit{not post--dislike}. If an agent performs multiple actions simultaneously at a given timestep, the Shapley value attribution is proportionally distributed across the performed actions.

In Figure~\ref{fig:action}(c), actions are indexed according to this combination of posting and interaction decisions. Specifically, actions 1--3 correspond to agents choosing not to post, with interactions \textit{like}, \textit{not interact}, and \textit{dislike}, respectively. Actions 4--6 correspond to agents choosing to post, with the same three interaction options in the order \textit{like}, \textit{not interact}, and \textit{dislike}.

In counterfactual simulations, each action is replaced by a predefined baseline behavior. In SocialNetwork, the baseline action is defined as taking no action at the current timestep, meaning the agent neither posts content nor interacts with any posts. All actions in counterfactual trajectories are substituted with this baseline for Shapley value computation.

\subsubsection{Belief Update Mechanism}

Agent beliefs evolve through two primary mechanisms: \textit{interaction-based updates} and \textit{feedback-based updates}. 

\textbf{Interaction-based updates} occur when an agent engages with content posted by others. When an agent $i$ with belief $b_i^t$ interacts with a post containing belief value $b_p$, the belief update depends on the interaction type:
\begin{equation}
    b_i^{t+1} = \begin{cases}
        b_i^t + \delta \cdot s_i^t \cdot |b_p - b_i^t| & \text{if } \text{like} \text{ and } b_p > b_i^t \\
        b_i^t - \delta \cdot s_i^t \cdot |b_p - b_i^t| & \text{if } \text{like} \text{ and } b_p \leq b_i^t \\
        b_i^t - \delta \cdot s_i^t \cdot |b_p - b_i^t| & \text{if } \text{dislike} \text{ and } b_p > b_i^t \\
        b_i^t + \delta \cdot s_i^t \cdot |b_p - b_i^t| & \text{if } \text{dislike} \text{ and } b_p \leq b_i^t
    \end{cases}
\end{equation}
where $\delta$ is the update magnitude parameter, and $s_i^t$ is the belief update sensitivity of agent $i$ at time $t$, defined as:
\begin{equation}
    s_i^t = s_{\text{base}} \cdot (1 - \alpha \cdot |b_i^t|),
\end{equation}
with $s_{\text{base}}$ being the base sensitivity and $\alpha$ a sensitivity coefficient. This formulation ensures that agents with more extreme beliefs (higher $|b_i^t|$) exhibit reduced sensitivity to new information, reflecting the phenomenon of confirmation bias and belief rigidity.

\textbf{Feedback-based updates} occur when an agent receives engagement on their own posts. After agent $i$ publishes a post at timestep $t$, if the post receives at least one view ($V_i^t > 0$), the belief is updated based on the net engagement received:
\begin{equation}
    b_i^{t+1} = b_i^t \cdot \left(1 + \frac{L_i^t - D_i^t}{V_i^t} \cdot \rho\right),
\end{equation}
where $L_i^t$, $D_i^t$, and $V_i^t$ denote the number of likes, dislikes, and total views received by agent $i$'s post at timestep $t$, respectively, and $\rho$ is the reinforcement coefficient. This mechanism models social reinforcement, where positive feedback (net likes) strengthens the agent's current belief, while negative feedback (net dislikes) weakens it. If no views are received ($V_i^t = 0$), the belief remains unchanged.

Additionally, agent behavioral parameters---posting preference and interaction preference---are dynamically updated based on the agent's current belief magnitude:
\begin{align}
    p_i^t &= p_{\text{base}} + \beta \cdot |b_i^t|, \\
    q_i^t &= q_{\text{base}} + \gamma \cdot |b_i^t|,
\end{align}
where $p_i^t$ and $q_i^t$ represent the posting and interaction preferences, respectively, $p_{\text{base}}$ and $q_{\text{base}}$ are base preference values, and $\beta$ and $\gamma$ are preference coefficients. This design captures the observation that agents with more extreme beliefs tend to be more active in both content creation and engagement.

System risk is quantified as the level of belief polarization in the network, measured by the variance of agent beliefs at time $t$:
\begin{equation}
    \hat{R}_t(\tau) = \frac{1}{N} \sum_{i=1}^N (b_i^t - \bar{b}^t)^2,
\end{equation}
where $b_i^t$ represents the belief value of agent $i$ at time $t$, $\bar{b}^t$ is the mean belief across the population, and $N$ is the total number of agents. Extreme events are defined based on this risk indicator, with the threshold determined through expert annotation to reflect a level of polarization considered socially significant or extreme.

\section{Implementation Details of Baseline Methods}\label{app:baselines}

\subsection{Failure Taxonomy (FT)}\label{app:ft}

The \textbf{Failure Taxonomy (FT)} method proposed by Cemri et al.~\cite{cemri2025multi} introduces a structured framework for analyzing failures in multi-agent LLM systems. 
FT defines 14 fine-grained failure modes grouped into three high-level categories: 
(i) \textit{System Design Issues}, 
(ii) \textit{Inter-Agent Misalignment}, and 
(iii) \textit{Task Verification}. 
This taxonomy provides a principled vocabulary for characterizing why and how coordination among LLM-based agents breaks down.

In the original work, FT is applied as an annotation framework in which a Large Language Model (LLM) serves as a judge to label execution traces from multiple multi-agent benchmarks with one or more failure modes. 
The authors demonstrate that LLM-based annotations achieve high agreement with human expert judgments, supporting the reliability of the proposed taxonomy.

In our experiments, we adopt the failure mode definitions and scoring rubric introduced in~\cite{cemri2025multi} and use them as a baseline for quantitative behavior analysis. 
Rather than directly reproducing the original trace-level labeling pipeline, we pair the FT scoring table with a newly designed prompt that guides an LLM to score agent action trajectories under each failure category. 
This enables us to obtain a structured, quantitative assessment of agent behaviors through the lens of the FT framework, which we use for comparison with our proposed attribution method. 

The prompt used in our study is shown below.

\begin{lstlisting}[style=promptstyle, caption={FT Prompt for Econagent}, label={lst:ft_prompt_econagent}]
To help you understand multi-agent system behaviors and potential issues,
here are definitions of common failure modes in multi-agent systems:
{definitions}
Here are some examples of failure modes in multi-agent systems for reference:
{examples}
You are analyzing an economic simulation system. Below is a PARTIAL table of agent actions.
IMPORTANT: This is batch {} of {}.
Timestep range for this batch: {} to {}.
Target risk timesteps: {}
Analyze each action's contribution to the risk value.
Here is the action table for this batch:{}

Context information:
- Target timestep risk value: {:.6f}
- Baseline risk value: {:.6f}
- Risk difference: {:.6f}
- Number of agents: {}
- Number of timesteps: {}

Task: Score each action's contribution to the risk value on a scale of 0 to 1, where:
- 0 means the action has minimal/no contribution to the risk
- 1 means the action has maximum contribution to the risk
- Scores should reflect how much each action contributes to moving the risk from baseline to the target risk value
- Consider the action's impact on system behavior, coordination, and potential failure modes
- IMPORTANT: Only score the (agent_id, timestep) pairs that appear in the table above.
Please output your result in the following JSON format:
{
  "scores": [
    {
      "agent_id": <agent_id>,
      "timestep": <timestep>,
      "score": <float between 0 and 1>
    },
    ...
  ]
}
Ensure that:
- All agent-timestep pairs in THIS BATCH are covered.
- Output ONLY the JSON object, no additional text before or after.
\end{lstlisting}

\begin{lstlisting}[style=promptstyle, caption={FT Prompt for TwinMarket}, label={lst:ft_prompt_twinmarket}]
To help you understand multi-agent system behaviors and potential issues, 
here are definitions of common failure modes in multi-agent systems:
{definitions}
Here are some examples of failure modes in multi-agent systems for reference:
{examples}
You are analyzing a stock market simulation system. Below is a PARTIAL table of user trading actions.
IMPORTANT: This is batch {} of {}.
Date range for this batch: {} to {}.
Target risk date: {}
Analyze each action's contribution to the risk value.
Here is the action table for this batch:{}

Context information:
- Target date risk value: {:.6f}
- Baseline risk value: {:.6f}
- Risk difference: {:.6f}
- Number of users: {}
- Number of dates: {}

Task: Score each action's contribution to the risk value on a scale of 0 to 1, where:
- 0 means the action has minimal/no contribution to the risk
- 1 means the action has maximum contribution to the risk
- Scores should reflect how much each action contributes to moving the risk from baseline to the target risk value
- Consider the action's impact on market behavior, coordination, and potential failure modes
- IMPORTANT: Only score the (user_id, date) pairs that appear in the table above.
Please output your result in the following JSON format:
{
  "scores": [
    {
      "user_id": "<user_id>",
      "date": "<date>",
      "score": <float between 0 and 1>
    },
    ...
  ]
}
Ensure that:
- All user-date pairs in THIS BATCH are covered.
- Output ONLY the JSON object, no additional text before or after.
\end{lstlisting}

\begin{lstlisting}[style=promptstyle, caption={FT Prompt for SocialAgent}, label={lst:ft_prompt_socialagent}]
To help you understand multi-agent system behaviors and potential issues, 
here are definitions of common failure modes in multi-agent systems:
{definitions}
Here are some examples of failure modes in multi-agent systems for reference:
{examples}
You are analyzing a social media simulation system. Below is a table of agent actions at each timestep.
Analyze each action's contribution to the polarization risk value.
Here is the action table:
{action_table}

Context information:
- PEAK risk value at target timestep {target_timestep}: {real_risk:.6f}
- Initial baseline risk value: {baseline_risk:.6f}
- Risk increase from baseline to peak: {risk_diff:.6f}
- Number of agents: {n_agents}
- Number of timesteps: {n_timesteps}
- Target timestep (where peak risk occurs): {target_timestep}

Task: Score each action's contribution to reaching the PEAK risk value on a scale of 0 to 1, where:
- 0 means the action has minimal/no contribution to reaching the peak risk
- 1 means the action has maximum contribution to reaching the peak risk
- Scores should reflect how much each action contributes to moving the risk from baseline to the PEAK risk value at timestep {target_timestep}
- Consider the action's impact on polarization, echo chambers, and potential failure modes
Please output your result in the following JSON format:
{
  "scores": [
    {
      "user_id": "<user_id>",
      "timesteps": "<date>",
      "score": <float between 0 and 1>
    },
    ...
  ]
}
Ensure that:
- All user-date pairs in THIS BATCH are covered.
- Output ONLY the JSON object, no additional text before or after.
\end{lstlisting}

\subsection{Failure Attribution (FA)}\label{app:fa}

The \textbf{Failure Attribution (FA)} method proposed by Zhang et al.~\cite{zhang2025agent} aims to automatically identify the agent and timestep responsible for the failure of a multi-agent system. 
FA adopts a direct prompting strategy in which a Large Language Model (LLM) is provided with the complete execution trajectory of a failed multi-agent interaction and is asked to output the \emph{failure-responsible agent} together with the \emph{decisive error step}. 
Formally, a decisive error is defined as the earliest agent--step pair whose correction would change the system outcome from failure to success, following a counterfactual intervention on the original trajectory.

In the original formulation, FA is designed to produce a \emph{single} agent--step pair as the root cause of failure. 
The prompt therefore elicits a point prediction rather than a distribution over all agent actions, reflecting its focus on identifying one decisive causal mistake instead of quantifying the relative importance of multiple actions along the trajectory.

In our setting, however, the objective differs from identifying a unique decisive error in the strict counterfactual sense. 
Our goal is to attribute the observed system-level risk at a target timestep to prior agent actions by estimating their \emph{relative contributions} to risk escalation. 
To align FA with this objective while preserving its core intuition, we adapt the original FA prompt to accept our trajectory representation and instruct the LLM to output a score for each agent--step pair. 
A higher score indicates a greater contribution of action $a_{i,t}$ to the system risk at the target timestep, rather than serving as a binary indicator of a decisive error.

This adapted version of FA is used as a baseline attribution method in our experiments. 
While it departs from the original FA objective by producing a dense attribution over all agent actions, it retains the fundamental idea of leveraging LLMs to reason over complete multi-agent trajectories for failure-related analysis.

\begin{lstlisting}[style=promptstyle, caption={FA Prompt for EconAgent}, label={lst:fa_prompt_econagent}]
You are an AI assistant tasked with analyzing a multi-agent macroeconomic simulation trajectory.
IMPORTANT: This is batch {} of {}.
Timestep range for this batch: {} to {}.
Target risk timesteps: {}
The simulation consists of {} agents interacting over {} time steps.
At each time step, each agent makes economic decisions such as consumption and labor participation,
which jointly influence the macroeconomic price level and inflation dynamics.
The problem is:
The economic system exhibits a risk indicator that is significantly higher than a baseline system. 
The current system risk is {:.6f}, while the baseline risk is {:.6f}. 
The risk difference is {:.6f}.
The risk indicator is defined based on Engle (1982) and Bollerslev (1986) and measures the conditional variance
of inflation forecast errors, reflecting systemic economic risk.
Risk computation:
1. Inflation rate: $\pi_t = \log(P_t) - \log(P_{t-1})$ \\
2. Expected inflation (naive expectation): $E_{t-1}[\pi_t] = \pi_{t-1}$ \\
3. Forecast error: $e_t = \pi_t - E_{t-1}[\pi_t]$ \\
4. Risk indicator: $h_t = \lambda \cdot h_{t-1} + (1 - \lambda) \cdot e_{t-1}^2$, where $\lambda = {}$
Ground truth:
The baseline system risk trajectory h_t^baseline is provided as the reference.
The current system produces a much higher risk level than the baseline.
Your task:
Analyze the system trajectory and attribute the excessive risk to individual agents' actions
across time.

You are given a PARTIAL system execution trace (batch {} of {}) in chronological order.
Each entry represents one agent's action at a specific timestep, with the following fields:
- agent_id: The identifier of the agent (0 to {})
- timestep: The time step (1 to {})
- wealth: The agent's wealth at this timestep
- income: The agent's income at this timestep
- consumption_rate: The agent's consumption rate (0-1)
- job: The agent's job type

Here is the system trajectory for this batch:
{}
Based on this trajectory, please predict the following:
For EACH agent at EACH timestep in THIS BATCH ONLY, assign a risk attribution score in the range [0, 1],
indicating how much that specific agent's action at that timestep contributes to
the elevated system risk relative to the baseline.
Scoring principles:
- A higher score indicates a stronger causal contribution to increased inflation volatility
  and higher conditional variance h_t.
- Scores should reflect both direct and indirect effects of agents' actions on macroeconomic instability.
- If an action is largely irrelevant to the increased risk, assign a score close to 0.
- If an action is a major driver of risk escalation, assign a score close to 1.
- Scores should be comparable across agents and timesteps.
- Consider temporal causality: earlier actions can influence later risk, so actions at earlier timesteps
  may have higher scores if they trigger cascading effects.
- IMPORTANT: Only score the (agent_id, timestep) pairs that appear in the trajectory above.
Please output your result in the following JSON format:
{}
Ensure that:
- All agent-timestep pairs in THIS BATCH are covered.
- The attribution reflects temporal causality (earlier actions can influence later risk).
{}
- Output ONLY the JSON array, no additional text before or after.
\end{lstlisting}

\begin{lstlisting}[style=promptstyle, caption={FA Prompt for TwinMarket}, label={lst:fa_prompt_twinamrket}]
You are an AI assistant tasked with analyzing a stock market simulation trajectory.
The simulation consists of {} users trading stocks over {} days.
At each day, each user makes trading decisions (buy/sell orders),
which jointly influence the market price dynamics and systemic risk.
The problem is:
The market system exhibits a risk indicator that is significantly higher than a baseline system. 
The current system risk is {:.6f}, while the baseline risk is {:.6f}. 
The risk difference is {:.6f}.
The risk indicator measures the conditional variance of inflation forecast errors,
reflecting systemic economic risk based on market price volatility.
Risk computation:
1. Market average price is calculated from stock prices
2. Log return: $\pi_t = \log(P_t) - \log(P_{t-1})$
3. Expected return (naive expectation): $E_{t-1}[\pi_t] = \pi_{t-1}$
4. Forecast error: $e_t = \pi_t - E_{t-1}[\pi_t]$
5. Risk indicator: $h_t = \lambda \cdot h_{t-1} + (1 - \lambda) \cdot e_{t-1}^2$, where $\lambda = 0.94$
Ground truth:
The baseline system risk trajectory h_t^baseline is provided as the reference.
The current system produces a much higher risk level than the baseline.
Your task:
Analyze the system trajectory and attribute the excessive risk to individual users' actions
across time.

You are given the system execution trace in chronological order.
Each entry represents one user's trading activity at a specific date, with the following fields:
- user_id: The identifier of the user
- date: The trading date (YYYY-MM-DD)
- n_transactions: Total number of buy and sell sub-orders
- n_buy: Number of buy sub-orders
- n_sell: Number of sell sub-orders
- buy_amount: Total buy quantity * price
- sell_amount: Total sell quantity * price
- n_buy_stocks: Number of unique stocks bought
- n_sell_stocks: Number of unique stocks sold
- target_date: The date when risk occurred

Here is the system trajectory:
{}
Based on this trajectory, please predict the following:
For EACH user at EACH date, assign a risk attribution score in the range [0, 1],
indicating how much that specific user's trading actions at that date contribute to
the elevated system risk relative to the baseline.
Scoring principles:
- A higher score indicates a stronger causal contribution to increased price volatility
  and higher conditional variance h_t.
- Scores should reflect both direct and indirect effects of users' trading actions on market instability.
- If an action is largely irrelevant to the increased risk, assign a score close to 0.
- If an action is a major driver of risk escalation, assign a score close to 1.
- Scores should be comparable across users and dates.
- Consider temporal causality: earlier actions can influence later risk, so actions at earlier dates
  may have higher scores if they trigger cascading effects.
Please output your result in the following JSON format:
{}
Ensure that:
- All user-date pairs in the trajectory are covered.
- The attribution reflects temporal causality (earlier actions can influence later risk).
{}
- Output ONLY the JSON array, no additional text before or after.
\end{lstlisting}

\begin{lstlisting}[style=promptstyle, caption={FA Prompt for SocialAgent}, label={lst:fa_prompt_social}]
You are an AI assistant tasked with analyzing a social media simulation trajectory.
The simulation consists of {} agents interacting over {} timesteps.
At each timestep, agents can post content and interact with others' posts (view, like, dislike),
which jointly influence the polarization dynamics and systemic risk.
The problem is:
The social media system exhibits a polarization risk that is significantly higher than a baseline system. 
The current system risk is {:.6f}, while the baseline risk is {:.6f}. 
The risk difference is {:.6f}.
The polarization risk measures the variance and extremism of agent beliefs,
reflecting how polarized the social media ecosystem has become.
Risk computation:
1. Belief values range from -1 (extreme negative) to +1 (extreme positive)
2. Polarization risk = variance of beliefs + extremism penalty
3. Extremism penalty increases when beliefs are close to -1 or +1
Ground truth:
The baseline system risk trajectory is provided as the reference.
The current system produces a much higher risk level than the baseline.
Your task:
Analyze the system trajectory and attribute the excessive risk to individual agents' actions
across time.

You are given the system execution trace in chronological order.
Each entry represents one agent's activity at a specific timestep, with the following fields:
- agent_id: The identifier of the agent
- timestep: The timestep (0-indexed)
- posted: Whether the agent posted content (1) or not (0)
- view_count: Number of posts the agent viewed
- like_count: Number of posts the agent liked
- dislike_count: Number of posts the agent disliked

Here is the system trajectory:
{}
Based on this trajectory, please predict the following:
For EACH agent at EACH timestep, assign a risk attribution score in the range [0, 1],
indicating how much that specific agent's actions at that timestep contribute to
reaching the peak risk value ({:.6f}) at the target timestep ({}).
Scoring principles:
- A higher score indicates a stronger causal contribution to increased polarization
  and higher risk.
- Scores should reflect both direct and indirect effects of agents' actions on system instability.
- If an action is largely irrelevant to the increased risk, assign a score close to 0.
- If an action is a major driver of risk escalation, assign a score close to 1.
- Scores should be comparable across agents and timesteps.
- Consider temporal causality: earlier actions can influence later risk, so actions at earlier timesteps
  may have higher scores if they trigger cascading effects.
Please output your result in the following JSON format:
{}
Ensure that:
- All agent-timestep pairs in the trajectory are covered.
- The attribution reflects temporal causality (earlier actions can influence later risk).
{}
- Output ONLY the JSON array, no additional text before or after.
\end{lstlisting}

\subsection{Agent Tracer (AT)}\label{app:at}

The \textbf{Agent Tracer (AT)} method~\cite{zhang2025agentracer} adopts a supervised learning approach. 
It trains a neural network to estimate agent contributions by mapping state-action pairs to event outcomes. 
Specifically, AT leverages counterfactual replay and programmatic fault injection to construct a curated dataset of failure trajectories, where each trajectory is annotated with the decisive error step---the earliest agent--step pair whose correction would transform the system outcome from failure to success. 
The method trains a neural network to identify the decisive agent--step pair responsible for system failure, using counterfactual replay and curated failure trajectories annotated with the true error step.

In the original formulation, AT is designed to identify a \emph{single} decisive error step within a failed trajectory, focusing on pinpointing the root cause of system failure rather than quantifying the relative importance of multiple actions. 
The training process emphasizes both agent-level and step-level attribution accuracy through a reward function that combines format compliance, agent identification correctness, and temporal proximity to the true error step.

In our setting, however, the objective differs from identifying a unique decisive error in the strict counterfactual sense. 
Our goal is to attribute the observed system-level risk at a target timestep to prior agent actions by estimating their \emph{relative contributions} to risk escalation. 
To align with our objective while retaining the counterfactual attribution intuition emphasized by Agent Tracer, 
we construct an \textbf{AT-inspired baseline} that replaces the original learning-based surrogate model with a \textbf{Leave-One-Out (LOO)} counterfactual evaluation.

Specifically, for each agent--step pair $a_{i,t}$ in the trajectory, we compute its importance score as the risk reduction achieved by removing that action from the system. 
Formally, let $R_T(\tau)$ denote the system risk at the target timestep under the full trajectory $\tau$, and let $R_T(\tau_{-i,t})$ denote the risk when action $a_{i,t}$ is replaced with a baseline action. 
The LOO importance score for action $a_{i,t}$ is then defined as:
\begin{equation}
\text{LOO}(\text{agent}_i, t) = R_T(\tau) - R_T(\tau_{-i,t}),
\end{equation}
where a higher score indicates a greater contribution of the action to the system risk at the target timestep. 
This approach eliminates the need for model training while providing a principled, interpretable method for quantifying action importance through direct counterfactual intervention.

This AT-inspired LOO attribution method is used as a baseline in our experiments.
While it departs from the original AT objective by producing a dense attribution over all agent actions rather than identifying a single decisive error, it retains the fundamental idea of leveraging counterfactual reasoning to assess the causal impact of individual actions on system outcomes.


\end{document}